\RequirePackage{amsmath}
\documentclass[runningheads]{llncs}
\usepackage{siunitx}
\usepackage{cite}
\usepackage{epsfig}
\usepackage{graphicx}
\usepackage{amsmath}
\usepackage{dsfont}
\usepackage{textcomp}
\usepackage{boldline,multirow}
\usepackage[colorlinks=true,citecolor=blue,urlcolor=black]{hyperref}
\usepackage{url}
\usepackage{bbding}

\begin{document}
\title{Encoding CT Anatomy Knowledge for Unpaired Chest X-ray Image Decomposition}
\titlerunning{Encoding CT Anatomy Knowledge for Chest X-ray Decomposition}
%

\newcommand*\samethanks[1][\value{footnote}]{\footnotemark[#1]}

\author{Zeju Li\inst{1}\thanks{This work was done when Z. Li, G. Shi and J. Wang were interns at MIRACLE.} \and
Han Li\inst{2} \and
Hu Han\inst{3,4}\Envelope \and
Gonglei Shi \inst{5}\samethanks[1] \and
Jiannan Wang\inst{6}\samethanks[1] \and \\
S. Kevin Zhou\inst{3,4}\Envelope}

\institute{Biomedical Image Analysis Group, Imperial College London, London, UK \email{zeju.li18@imperial.ac.ukk} \and
University of Chinese Academy of Sciences, Beijing, China \and
Medical Imaging, Robotics, Analytic Computing Laboratory/Engineering (MIRACLE), Key Laboratory of Intelligent Information Processing of Chinese Academy of Sciences (CAS),
Institute of Computing Technology, CAS, Beijing, China \email{\{hanhu, zhoushaohua\}@ict.ac.cn} \and
Peng Cheng Laboratory, Shenzhen, China \and
Department of Computer Science and Technology, Southest University, Nanjing, China \and
College of Electronic Information and Optical Engineering, Nankai University, Tianjin, China}
%


\authorrunning{Zeju~Li et al.}
\maketitle              
\begin{abstract}
Although chest X-ray (CXR) offers a 2D projection with overlapped anatomies, it is widely used for clinical diagnosis. There is clinical evidence supporting that decomposing an X-ray image into different components (e.g., bone, lung and soft tissue) improves diagnostic value. We hereby propose a decomposition generative adversarial network (DecGAN) to anatomically decompose a CXR image but {\bf with unpaired data}. We leverage the anatomy knowledge embedded in CT, which features a 3D volume with clearly visible anatomies. Our key idea is to embed CT priori decomposition knowledge into the latent space of unpaired CXR autoencoder. Specifically, we train DecGAN with a decomposition loss, adversarial losses, cycle-consistency losses and a mask loss to guarantee that the decomposed results of the latent space preserve realistic body structures. Extensive experiments demonstrate that DecGAN provides superior unsupervised CXR bone suppression results and the feasibility of modulating CXR components by latent space disentanglement. Furthermore, we illustrate the diagnostic value of DecGAN and demonstrate that it outperforms the state-of-the-art approaches in terms of predicting 11 out of 14 common lung diseases.

\end{abstract}
\section{Introduction}

Chest X-ray (CXR) and computed tomography (CT) are two closely related medical imaging modalities given that a 3D CT is reconstructed from a set of X-ray projections. However, CXR is only a 2D projection image which contains overlapped anatomies and ambiguous structure details. A possible connection between CXR and CT is via a digitally reconstructed radiograph (DRR), which is a virtual CXR-like projection image calculated from a CT volume based on ray-tracing and rigid transform. The decomposition of DRR is readily accessible.

\begin{figure}[t]
  \centering
  \includegraphics[width=\textwidth]{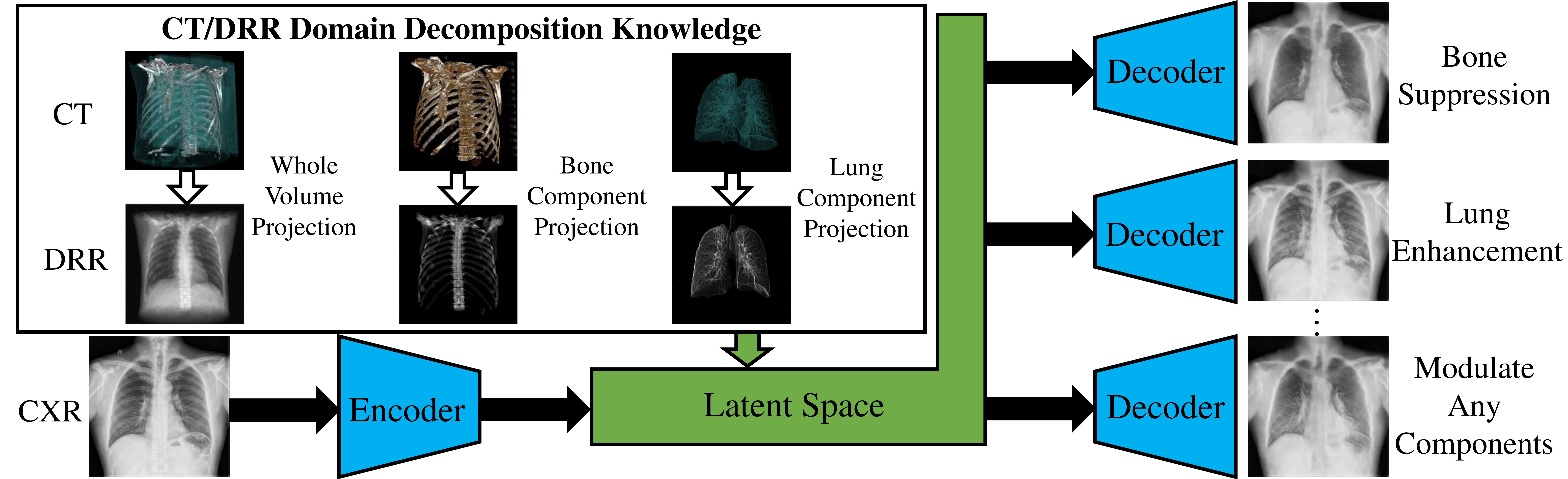}
  \caption{We aim to decompose a chest X-ray (CXR) into multiple components by utilizing priori decomposition knowledge
  provided by computed  tomography (CT) domain.}
 \label{fig:Fig1_MainIdea}
  \end{figure}

In terms of economy and health, it could be of great importance to improve the diagnostic value of CXR. There is clinical evidence supporting that decomposing an X-ray image into different components (e.g. bone, lung and soft tissue) improves diagnostic value~\cite{39laskey1996dual}. This paper aims to propose a method to decompose a CXR into different components such as bone, lung and other soft-tissue structures, thereby boosting the CXR diagnosis accuracy of lung diseases. Existing approaches to X-ray image decomposition are supervised, requiring paired original and decomposed images in training. For example, the paired dual energy (DE) imaging is always needed for supervised bone suppression and serves as the learning targets of neural networks~\cite{17yang2017cascade, zhou2018generation}. DE radiography provides bone-free CXR by capturing two radiographs at two energy levels. However, these methods are limited because only few hospitals could provide DE images, and the obtained learning models might do not work well on other CXR datasets. We attempt to offer a solution \emph{without DE images}. As shown in Fig.~\ref{fig:Fig1_MainIdea}, the main idea of our proposed method, called a decomposition generative adversarial network (DecGAN), is to integrate the decomposition knowledge from the CT domain into the latent space of CXR autoencoder. The DRR decomposition is generated by projecting separate CT anatomic components obtained in a 3D CT volume.

DecGAN is developed upon the theory of domain adaptation based on generative adversarial network (GAN)~\cite{9zhu2017unpaired}. For medical imaging, it is more crucial to resolve the domain adaptation problem because medical data is always limited~\cite{23zhang2018task}. However, currently all these domain adaptation methods for medical imaging are designed for the tasks of image segmentation or classification but not for image decomposition, which is more difficult for domain adaptation because it is required to preserve a lot of detail information. To the best of our knowledge, DecGAN is {\em the first approach} to tackle the problem of cross-domain medical image decomposition under an unpaired setting. 

In this paper, we overcome three main challenges in cross-domain image decomposition: 1) creating a corresponding latent space, 2) building a powerful and robust decoder, and 3) preserving anatomical reliability. We demonstrate that our method outperforms other state-of-the-art methods on the task of unsupervised CXR bone suppression and allows modulating CXR components, which is of great clinical value. Our preliminary study shows that DecGAN can enhance the diagnostic potential of CXR and benefit the diagnosis of lung diseases.

\section{Method}

\begin{figure*}
\centering
\includegraphics[width=\textwidth]{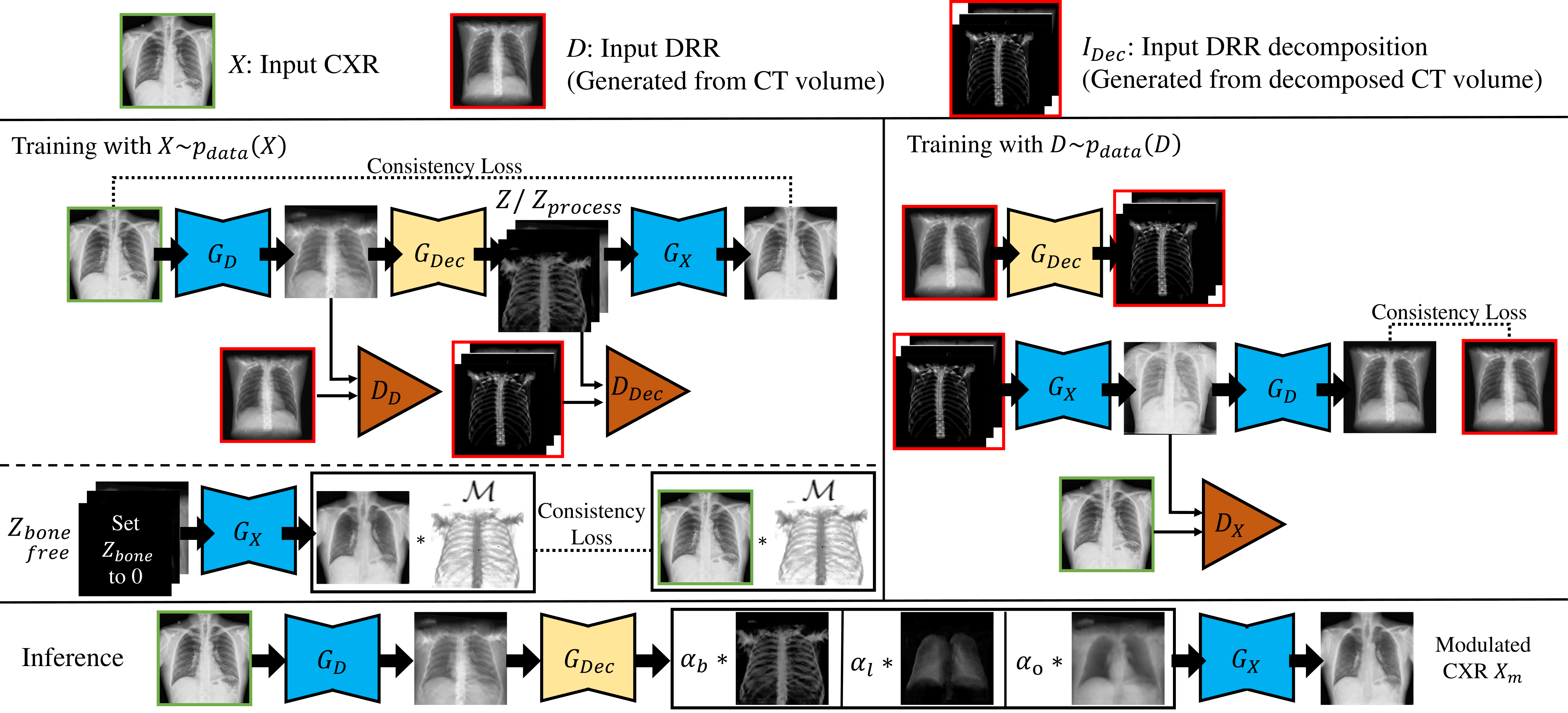} \caption{The illustration of the proposed method. \textbf{Training phase:} Decomposition network \emph{G$_{Dec}$} and decomposition discriminator network \emph{D$_{Dec}$} are embedded into the backbone of CycleGAN which is designed to connect the domains of CXR and digitally reconstructed radiograph (DRR). A mask loss is provided to guarantee the corresponding relationship of non-bone components between the latent space and the reconstructed images. \textbf{Inference phase: } The components of reconstructed CXR can be modulated by changing the weight of probability maps in the latent space.}
\label{fig:Fig2_ProposedMethod}
\end{figure*}

DecGAN is designed upon the backbone of CylceGAN with latent space disentanglement, as shown in Fig.~\ref{fig:Fig2_ProposedMethod}. Given a CXR \emph{X} as input, we want to build a function \emph{F} to produce the modulated reconstruction \emph{X$_{m}$}, in which different chest components can be modulated by changing the corresponding factors $[\alpha_{b},\alpha_{l},\alpha_{o}]$:

\begin{equation}
\emph{X$_{m}$}=F(X, \alpha_{b}, \alpha_{l}, \alpha_{o})=G_X(G_{Dec}(G_D(X, \alpha_{b}, \alpha_{l}, \alpha_{o}))), \label{eq:r0}
\end{equation}

In order to tackle the problem of CXR decomposition, we mainly make three contributions: 1) An additional latent space decmposition discriminator \emph{D$_{Dec}$} is designed to encourage the embedding of priori CT decomposition knowledge and the separation of different components in the generated DRR. 2) The DRR decomposition network \emph{G$_{Dec}$} is embedded into the backbone of CycleGAN to provide the decoder enough knowledge to tackle the decomposition information in the latent space. 3) A soft bone mask ${\cal M}$, generated using the bone components in the latent space, serves as additional constraints to make the components separated and realistic in the reconstruction results.

\subsection{DRR Decomposition Network G$_{Dec}$}
The decomposition network is designed based on the architecture of U-net~\cite{35ronneberger2015u}. The components of a 3D CT volume are projected using the same parameters and concatenated into channels to serve as the ground truths of DRR decomposition:
\begin{equation}
\emph{I$_{Dec}$}=[\emph{I$_{bone}$},\emph{I$_{lung}$},\emph{I$_{other}$}], \label{eq:r1}
\end{equation}
where \emph{I$_{bone}$}, \emph{I$_{lung}$} and \emph{I$_{other}$} are the components of DRR for bone, lung and other soft-tissue, respectively.

Based on the input DRR image \emph{D} and its separated components \emph{I$_{Dec}$}, the decomposition network can be trained in a supervised way by using the decomposition loss:
\begin{gather}
\mathcal{L}_{Dec}(G_{Dec})=\mathds{E}_{D_{\sim}p_{data}(D)}[\|G_{Dec}(D)-I_{Dec}\|_2^2]. 
\label{eq:r2}
\end{gather}

\subsection{Generative Models G$_{D}$ and G$_{X}$}

The generative models are trained in an unsupervised way based on unpaired CXR and DRR. The DRR generation network \emph{G$_{D}$} is trained to generate realistic DRR based on input CXR, while the CXR generation network \emph{G$_{X}$} is trained to reconstruct a realistic CXR based on the input DRR components.

\textbf{Training with CXR.}
Taking the CXR \emph{X} as input, the probability maps of decomposed components can be obtained using \emph{G$_{D}$} and \emph{G$_{Dec}$}:
\begin{equation}
Z=G_{Dec}(G_D(X))=[Z_{bone},Z_{lung},Z_{other}],\label{eq:r3}
\end{equation}
where \emph{Z$_{bone}$}, \emph{Z$_{lung}$} and \emph{Z$_{other}$} refer to the probability maps of bone, lung and other soft-tissue structures, respectively, which also correspond to (\ref{eq:r1}). Thereafter, these maps are encoded into \emph{Z$_{process}$} to ensure the information is complete in the latent space:
\begin{equation}
Z_{process}=[Z_{bone},Z_{lung},G_D(X)-Z_{bone}-Z_{lung}].\label{eq:r4}
\end{equation}

Once different CXR components are encoded separately into \emph{Z$_{process}$}, it is possible to highlight or suppress specific CXR components by modulating the weights of different components in \emph{Z$_{process}$}.

Another decomposition discriminator \emph{D$_{Dec}$} is added in DecGAN to help separate different CXR components in the latent space. Thus, the generation loss is given as:
\begin{equation}
\begin{aligned}
\mathcal{L}_{GXD}(G_D, D_D,D_{Dec})= \mathds{E}_{X_{\sim}p_{data}(X)}[log(1-D_D(G_D(X)))\\+ log(1-D_{Dec}(Z_{process}))]+\mathds{E}_{D_{\sim}p_{data}(D)}[log(D_D(D))+log(D_{Dec}(I_{Dec}))].
\end{aligned}\label{eq:r5}
\end{equation}

The discriminators are designed to reduce the gap between DRR and the generated CXR as well as the gap between DRR decomposition and CXR decomposition in the latent space.

At last, the cycle-consistency loss is applied to
constrain the reconstruction results:
\begin{align}
\mathcal{L}_{cycX}&(G_D,G_X)=\mathds{E}_{X_{\sim}p_{data}(X)}[\|G_X(Z_{process})-X\|_1].\label{eq:r6}
\end{align}

\textbf{Training with DRR.}
The training cycle of DRR is conventional except that the cycle of DRR starts from its decompositions instead of
the original images. The generation loss and cycle-consistency loss are defined as:
\begin{align}
\mathcal{L}_{GDX}&(G_X,D_X)=\mathds{E}_{X_{\sim}p_{data}(X)}[log(D_X(X)))\notag\\
+&\mathds{E}_{D_{\sim}p_{data}(D)}[log(1-D_X(G_X(I_{Dec})))],
\end{align}
\begin{align}
\mathcal{L}_{cycD}&(G_D,G_X)=\mathds{E}_{D_{\sim}p_{data}(D)}[\|G_D(G_X(I_{Dec}))-D\|_1].\label{eq:r8}
\end{align}

\textbf{Mask Loss.} 
In early experiments of DecGAN, we find that the reconstruction results lack fidelity when changing the probability maps \emph{Z$_{process}$}. It is because there is little priori knowledge for \emph{G$_{X}$} to generate results with changed \emph{Z$_{process}$}. To resolve this issue, we propose to put more constraints on the generative model for the CXR decomposition. Bone component is suitable for this task because it only appears in certain regions of CXR, and it is  distinguishable from other components. In addition, it is  always  of no use for lung disease diagnosis. Therefore, we decide to optimize the generative model by putting less `emphasis' on bone structures. 

Different from the complete probability maps \emph{Z$_{process}$} in (\ref{eq:r4}),
we first eliminate the bone components in the latent space:
\begin{equation}
Z_{bonefree}=[0,Z_{lung},G_D(X)-Z_{bone}-Z_{lung}].
\end{equation}

A soft mask ${\cal M}$ is then generated based on the existing bone probability map over 95\% confidence. As the images are normalized to [0,1], the soft mask is defined as:
\begin{equation}
{\cal M} =1-(Z_{bone}-t)/(1-t)*\delta[Z_{bone}],\label{eq:r12}
\end{equation}
where $t$ is a threshold (we set $t=0.95$) and the binary function \emph{$\delta[Z_{bone}]$} is defined as:
\begin{equation}
\delta[Z_{bone}]=\left\{
\begin{array}{rcl}
0       &      & {Z_{bone} <    t};\\
1       &      & {Z_{bone} \geq t }.
\end{array} \right.
\end{equation}

Based on the mask ${\cal M}$ in (\ref{eq:r12}), the reconstruction results without bone components or the bone suppression results can be restricted by the input CXRs. The mask loss drives the non-bone regions of the reconstruction results become more similar to the original images:

\begin{equation}
\mathcal{L}_{mask}(G_X)=\mathds{E}_{X_{\sim}p_{data}(X)}[\|G_X(Z_{bonefree})*{\mathcal M}-X*{\cal M}\|_1].\label{eq:r13}
\end{equation}

\subsection{Inference and Modulation}
In the inference phase, as the CXR components can be separated in the latent space and we have built the
generative model to decode this information, we can modulate the CXR components to highlight specific component (e.g., lung) by changing the weights of probability maps, $[\alpha_{b},\alpha_{l},\alpha_{o}]$ and generating the modulated CXR reconstruction like:

\begin{equation}
X_{m}=G_{X}([\alpha_{b}*Z_{bone},\alpha_{l}*Z_{lung},\alpha_{o}*(G_D(X)-Z_{bone}-Z_{lung})]). \label{eq:r14}
\end{equation}

\section{Experiments}

\begin{table}[t]
\centering
\caption{The quantitative results of unsupervised CXR bone suppression.}\label{tab1}
\newsavebox{\tablebox}
\begin{lrbox}{\tablebox}

\begin{tabular}{m{12mm}<{\centering}|m{8mm}<{\centering}|m{20mm}<{\centering}|m{25mm}<{\centering}|m{20mm}<{\centering}|m{13mm}<{\centering}|m{28mm}<{\centering}
}
\hlineB{3}
Method  &  CXR & Blind Signal Separation \cite{15hogeweg2013suppression} &  No Adaptation~\cite{35ronneberger2015u} & CycleGAN \cite{9zhu2017unpaired} & DecGAN  & DecGAN  
(no \emph{D$_{Dec}$}, no $\mathcal{L}${$_{mask}$)} 
\\
\hlineB{2}
\emph{r$_l$(10$^4$)} &  3.82  & 2.75 & 1.41 &  1.50  &  \textbf{0.854} & 1.03 
\\
PSNRs &  ------  & 28.2 & 26.7 &  27.1 & \textbf{29.6}  & 27.5 
\\
\hline
\end{tabular}
\end{lrbox}
\scalebox{0.911}{\usebox{\tablebox}}
\end{table}

\begin{figure}[t]
\centering
\includegraphics[width=0.72\textwidth]{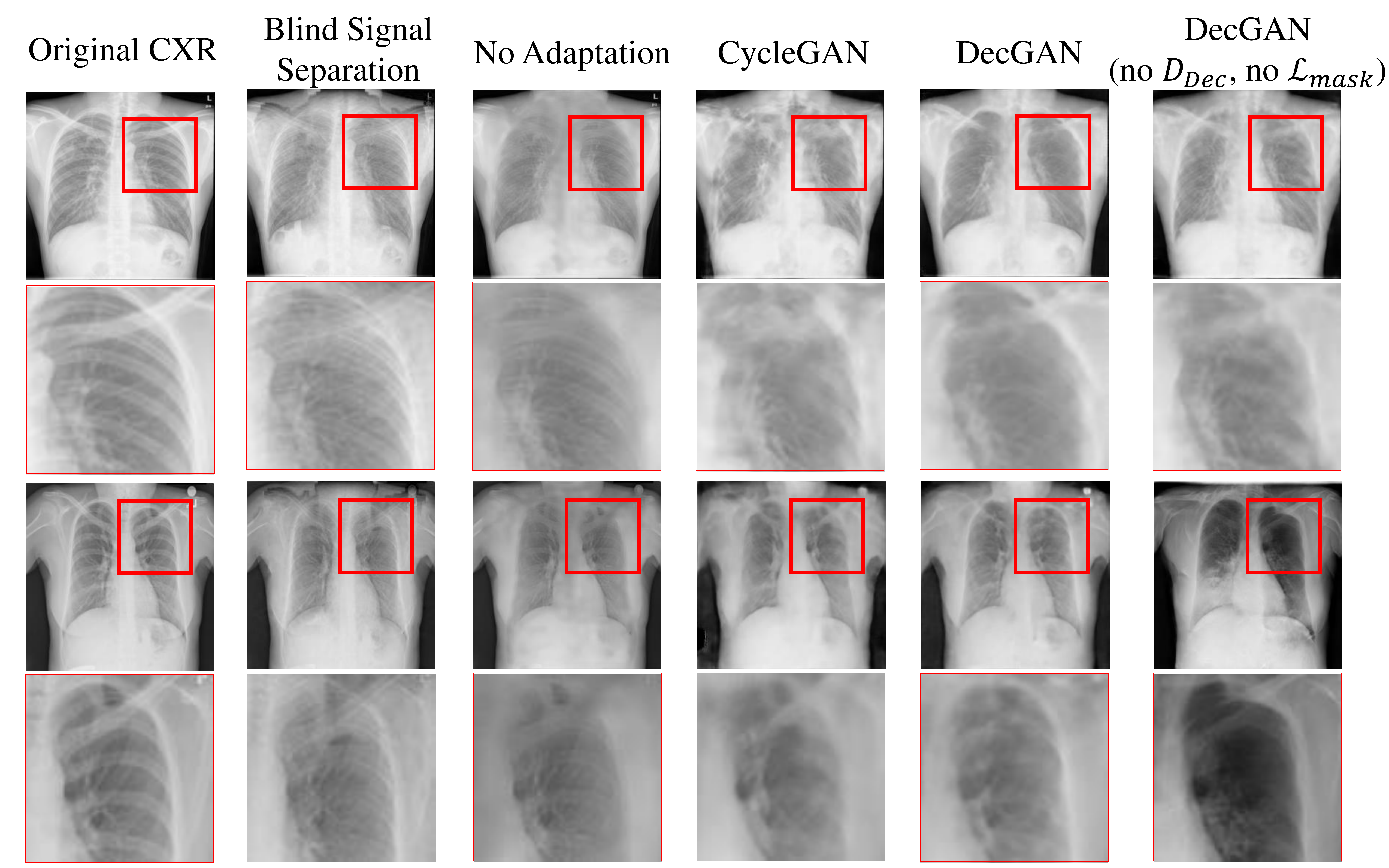}
\caption{Qualitative results of CXR bone suppression. We build several constraints in DecGAN to not only decompose different components separately but also produce realistic results. DecGAN can not only suppress bone to the greatest extent but also preserve the most realistic results of non-bone regions.}
\label{fig:Fig5_ResultsOnBoneSuppression}
\end{figure}

\subsection{Datasets}

We collect 246 CT volumes from LIDC-IDRI~\cite{10armato2011lung}. The bone regions in 3D CT volumes are labeled manually. The lung regions in CT volumes are labeled based on intensity and dilatation. We generate DRRs using those 246 CT volumes with augmentation based on rotation and rescaling. We collect 662 and 112,120 CXRs from Shenzhen Hospital X-ray Set~\cite{11jaeger2014two} and  ChestX-ray14~\cite{12wang2017chestx}, respectively. We randomly select 99 cases from Shenzhen Hospital X-ray Set as test data and use the official split of ChestX-ray14 in our experiments.

\subsection{CXR Bone Suppression}

We first assess the quality of CXR decomposition through the task of unsupervised bone suppression. We evaluate the results based on CXR from the Shenzhen Hospital X-ray Set because its superior image quality makes it easier for manual bone annotation. Like the previous studies~\cite{15hogeweg2013suppression}, we calculate the \emph{line response} $r_l$ to provide the quantitative evaluation of bone elimination, which is defined as the squared errors between the two eigenvalues of the Hessian matrix to reflect the elimination of high frequency bone borders. The lower is the metric, the less visible are the bone components in the CXRs. We also calculate \emph{peak signal to noise ratio} (PSNR) of the non-bone regions based on manual annotations to gauge how well the non-bone regions are preserved. The quantitative and qualitative results are summarized in Table~\ref{tab1} and Fig.~\ref{fig:Fig5_ResultsOnBoneSuppression}, respectively. DecGAN clearly outperforms the traditional solution which is based on blind signal suppression. Without domain adaptation, deep learning based methods would fail when the test CXR looks different from DRR: the bone suppression always lacks precision and the non-bone components are suppressed too. Similar failures happen to CycleGAN as the decoder has little knowledge of modified probabiliy maps, the non-bone regions of the reconstruction results tend to be disturbed and collapse. Integrating \emph{G$_{Dec}$} can make the generative model generate more reasonable bone suppression results, as shown in the results of `DecGAN (no \emph{D$_{Dec}$}, no $\mathcal{L}_{mask}$)'. However, the probability maps of different components are not separated clearly and the non-bone components sometimes look different from the original CXR. Therefore, the \emph{D$_{Dec}$} component is added to encourage the components to separate in the latent space and an additional mask loss based on bone regions is further designed for DecGAN to guarantee the components of the lung and other soft-tissue structures can remain unchanged through the generative model.

\subsection{Modulating CXR Components and Application to Diagnosis}

\begin{figure}[t]
\centering
\includegraphics[width=0.96\textwidth]{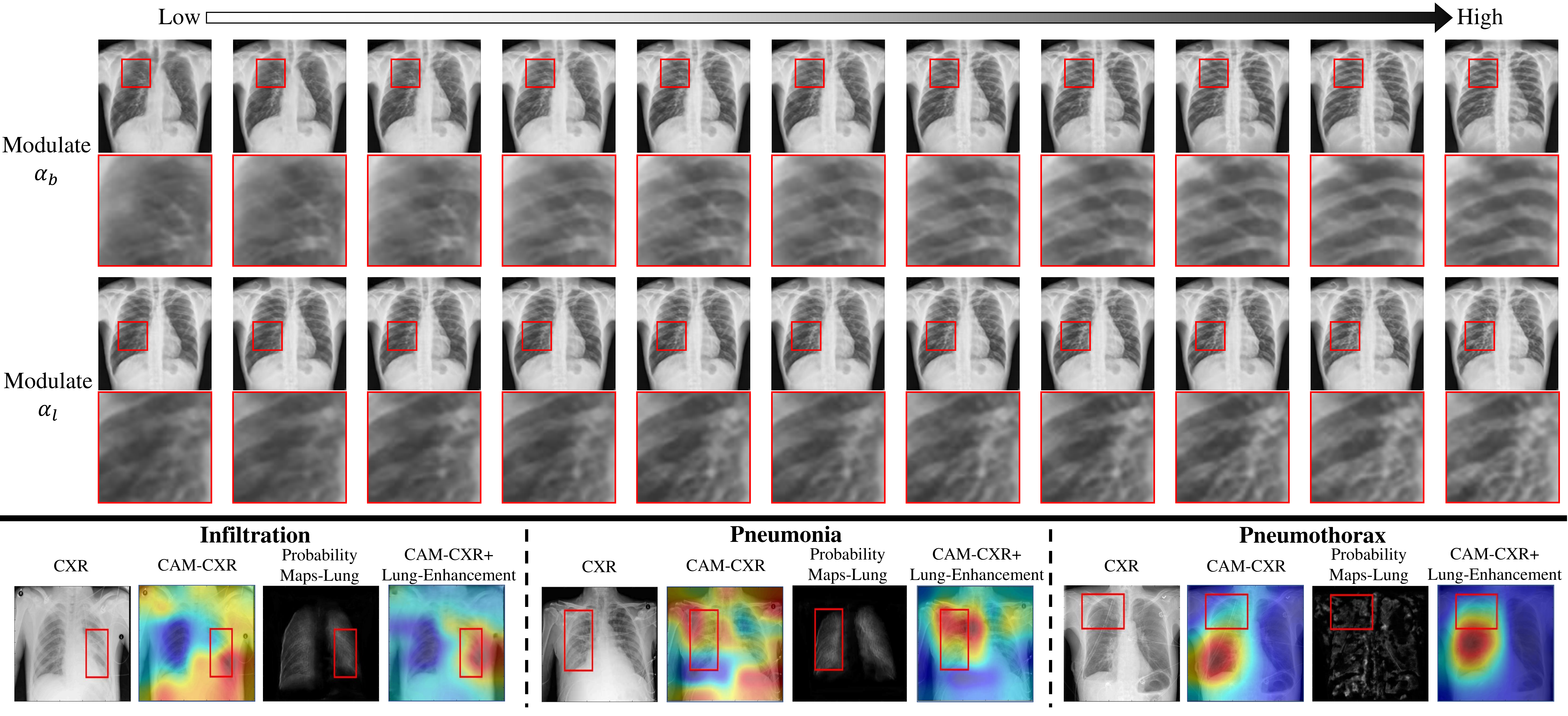}
\caption{(top half) Illustration of CXR modulation. DecGAN can modulate specific components with others unchanged. (bottom half) Examples of the unsupervised detection of lung diseases. The lung enhancement results can help better recognize and detect both the lung opacities, which are related to infiltration or pneumonia, and the lung collapse, which is caused by pneumothorax.}
\label{fig:Fig4_ModulateComponents_WeaklySupervisedResults}
\end{figure}

CXR components can be modulated by changing the weights of \emph{Z$_{process}$}, according to (\ref{eq:r14}). The results of modulating components are shown in Fig.~\ref{fig:Fig4_ModulateComponents_WeaklySupervisedResults}. The weights $\alpha${$_{b}$} and $\alpha${$_{l}$} are changed with the other weights always remaining the same. It is quite noticeable that the components of bone and lung are accordingly suppressed or enhanced while other components unchanged. This modulation feature has the potential of largely increasing the diagnosis value of CXR especially when the disease exists in a single component but overlaps with others.

In order to directly demonstrate the effectiveness of DecGAN for lung diseases diagnosis, the decomposition results are fed into the lung disease prediction system based on DenseNet-121. The lung enhancement results are generated with weights [$\alpha${$_{b}$}, $\alpha${$_{l}$}, $\alpha${$_{o}$}] being [1,2,1]. The lung enhancement results are concatenated with the original CXRs and stacked as the inputs of pretrained DenseNet-121~\cite{huang2017densely} model. The quantitative prediction results are summarized in Table~\ref{tab2}, along with others' results which are also evaluated based on the official splits. Our method achieves the state-of-the-art results on 11 out of 14 common lung diseases. While this performance boost is achieved \emph{under an unpaired training setting}, we believe that this boost will be more conspicuous with supervised training data. The class activation mapping (CAM) results are shown in Fig.~\ref{fig:Fig4_ModulateComponents_WeaklySupervisedResults}. The lung enhancement results by DecGAN are especially helpful in highlighting the lung abnormalities such as opacities in infiltration. In addition, the abnormal regions always show heterogeneity in the probability map of lung. For example, some disease patterns such as lung collapse in pneumothorax are more obvious in the probability map.

\section{Conclusion}

In this paper, we propose to learn a DecGAN \emph{under an unpaired setting}, which decomposes a CXR image into different components based on priori CT anatomy knowledge. We demonstrate the effectiveness of DecGAN in the tasks of unsupervised bone suppression and lung diseases diagnosis, achieving the state-of-the-art performances. We believe DecGAN can enhance the diagnostic potential of CXR.

\begin{table}[t]
\centering
\caption{The area under curve (AUC) of predicting 14 lung diseases from the ChestX-ray14 dataset. DecGAN can boost the prediction performance of the majority of lung diseases based on the prediction model of DenseNet-121.}\label{tab2}
\begin{lrbox}{\tablebox}
\begin{tabular}{c|c|c|c|c|c|c|c}
\hlineB{3}
Method  &  Atelectasis & Cardiomegaly  &  Effusion  & Infiltration & Mass & Nodule & Pneumonia \\
\hlineB{2}
Wang el al.\cite{12wang2017chestx} &  0.700  & 0.810 & 0.759 & 0.661 & 0.693 & 0.669 & 0.658  \\
DenseNet-121\cite{huang2017densely} &  0.777  & 0.879 & 0.825 &  0.696 & \textbf{0.836} & 0.773 & 0.730  \\
Ours &  \textbf{0.781}  & \textbf{0.881} & \textbf{0.827} &  \textbf{0.703} & 0.835 & \textbf{0.778} & \textbf{0.737} \\
\hlineB{3}
 Method& Pneumothorax & Consolidation & Edema & Emphysema & Fibrosis & Hernia&Pleura Thicken\\
 \hlineB{2}
Wang el al.\cite{12wang2017chestx}& 0.799 & 0.703 & 0.805 & 0.833 & 0.786 & 0.872&0.684 \\
DenseNet-121\cite{huang2017densely}& 0.842 & 0.761 & 0.847 & \textbf{0.920} & 0.823 & \textbf{0.938}& 0.779\\
Ours& \textbf{0.843} & \textbf{0.762} & \textbf{0.851} & 0.917 & \textbf{0.837}  & 0.929&\textbf{0.783} \\
\hline
\end{tabular}
\end{lrbox}
\scalebox{0.841}{\usebox{\tablebox}} 
\end{table}

\section*{Acknowledgements}
H. Han's work is supported in part by the Natural Science Foundation of China (grants 61732004 and 61672496), External Cooperation Program of Chinese Academy of Sciences (CAS) (grant GJHZ1843), and Youth Innovation Promotion Association CAS (grant 2018135).
We thank Cheng Ouyang for helpful comments.

%
%
%
%
\bibliographystyle{splncs04}
\bibliography{reference}

\end{document}